\documentclass[aps,prb,twocolumn,showpacs,superscriptaddress,10pt]{revtex4-1}
\usepackage{epsfig}
\usepackage{epstopdf}
\usepackage{graphicx}
\usepackage{bm}
\usepackage{amssymb}
\usepackage{amsmath}
\usepackage{amsthm}
\usepackage[utf8]{inputenc}

\newcommand{\BM}[1]{\mbox{\boldmath$#1$}}

\newcommand{\Br}{{\BM r}}

\newcommand{\Be}{{\BM e}}

\newcommand{\Bepsilon}{{\BM \epsilon}}

\newcommand{\BMM}{{\BM M}}

\usepackage{hyperref}

\begin{document}
\title{Disorder is good for you: The influence of local disorder on strain localization and ductility of strain softening materials}
\author{D\'aniel T\"uzes}
\email{tuzes@metal.elte.hu}
\affiliation{Institute for Materials Simulation (WW8), Friedrich-Alexander-University, Erlangen-Nürnberg, Dr.-Mack-Str. 77, D-90762 Fürth, Germany}
\affiliation{Department of Materials Physics, E\"otv\"os University, P\'azm\'any P\'eter s\'et\'any 1/a, H-1117 Budapest, Hungary}
\author{Michael Zaiser}
\affiliation{Institute for Materials Simulation (WW8), Friedrich-Alexander-University, Erlangen-Nürnberg, Dr.-Mack-Str. 77, D-90762 Fürth, Germany}
\author{P\'eter Dus\'an Isp\'anovity}
\affiliation{Department of Materials Physics, E\"otv\"os University, P\'azm\'any P\'eter s\'et\'any 1/a, H-1117 Budapest, Hungary}

\begin{abstract}
We formulate a generic concept model for the deformation of a locally disordered, macroscopically homogeneous material which undergoes irreversible 
strain softening during plastic deformation. We investigate the influence of the degree of microstructural heterogeneity and disorder on the 
concomitant strain localization process (formation of a macroscopic shear band). It is shown that increased microstructural heterogeneity delays 
strain localization and leads to an increase of the plastic regime in the macroscopic stress-strain curves. The evolving strain localization patterns are characterized and compared to models of shear band formation published in the literature. 
\end{abstract}

\maketitle

\section{Introduction}

Strain softening, loosely defined as a decrease of load carrying capability with increasing plastic deformation of a material, leads to strain localization (formation of shear bands) which in turn may lead to catastrophic failure of a material. If the width of the shear band is small as compared to the specimen dimensions, the macroscopic strain associated with the localized deformation may be small and failure occurs immediately after the material enters the softening regime. In materials where irreversible softening occurs shortly after yield, this may lead to a brittle appearance of the stress strain curves even though the failure mode is actually ductile. The most prominent example of this type of behavior are metallic glasses -- a class of materials with potentially outstanding mechanical properties \cite{Ashby2006} but whose application is hindered by a propensity to fail shortly after yield by catastrophic shear band formation. The softening mechanism is in this case most likely associated with a shear-induced increase in free volume \cite{Steif1982} though thermal softening associated with localized, adiabatic heating has been discussed as an alternative explanation (see e.g. \cite{Wright2001}).

Metallic glasses are an obvious example of materials which exhibit local structural disorder -- in this case down to the atomic scale. However, if one looks at defect microstructures, even crystalline solids exhibit (micro)structural disorder on scales well below the scale of a typical macroscopic specimen. On even larger scales, microstructural disorder is present in solid foams. In all these materials, one may legitimately ask how the macroscopic deformation behavior is influenced by the microstructural disorder and the associated length scales -- which in the examples given may range from nanometers (for metallic glasses) up to millimetres for solid foams. For the case of transient softening, as observed in compression of metallic foams, it has been shown that increasing the microstructural heterogeneity may actually lead to a more homogeneous distribution of deformation on the macroscopic scale \cite{Zaiser2013}. In the present paper we consider a generic model which accounts for heterogeneity and randomness in the material microstructure and microstructure evolution, in conjunction with strain softening. The model builds upon the scalar plasticity model of Zaiser and Moretti \cite{Zaiser2005} which was originally introduced for single-slip deformation of crystals with disordered dislocation microstructure, but has recently been used by many authors to model the inception of shear bands in amorphous materials and the associated avalanche phenomena (see e.g. \cite{Talamali2012, Budrikis2013,Sandfeld2015,Lin2015}). We generalize this model to explicitly introduce a strain softening mechanism. We first describe the model and then use it to study how the simulated deformation behavior depends on the degree of microstructural disorder (scatter of the distribution of local flow stresses). In particular we study the strain localization process and the concomitant stress strain curves, which demonstrate that increasing the disorder can delay strain localization and thus lead to a significant increase in macroscopic ductility.

\section{The stochastic continuum plasticity model}
The model was originally formulated for single slip crystal plasticity. Accordingly, the plastic strain is characterized by a scalar shear strain variable $\gamma$. Plastic deformation is assumed to proceed in discrete, localized events which occur once the local stress in a volume element exceeds a threshold value. An elementary slip event at $\Br$ creates a localized plastic Eigenstrain $\Delta\Bepsilon^{\rm pl}(\Br) = \Delta\gamma^{\rm pl} \BMM \delta(\Br)$ which we model as a point-like Eshelby inclusion \cite{Eshelby1957}. As a consequence of such an event, the internal stress field in the specimen volume changes. Increases in local stress in parts of the volume may trigger further localized events, leading to an avalanche which only terminates once the local stresses in all volume elements fall below the respective thresholds. 

In the present work we consider a 2D system where an infinitely extended specimen mimicked by periodic boundary conditions is, by remote boundary displacements, subject to a pure shear stress in the $xy$ plane. Thus, the stress tensor $\mathbf{\sigma}$ has only one independent component $\tau \left( {\Br} \right) := {\sigma _{xy}}\left( {\Br} \right)$. As stated above we assume that also the plastic strain tensor has only one independent component, hence  ${\Bepsilon^{\rm pl}}(\Br) = \gamma^{\rm pl}\left( {\Br} \right)\BMM$ where $\gamma^{\rm pl}\left( {\Br} \right)$ is the local plastic strain field and $\BMM = \left({\Be_{y}} \otimes {\Be_{x}} + {\Be_{x}} \otimes {\Be_{y}}\right)/2$. The stress acting on a volume element at $\Br$ can then be evaluated as the sum of the external stress and the internal stress associated with the inhomogeneous plastic strain field, ${\tau ^{{\rm loc}}}\left( {\Br} \right) = {\tau ^{\operatorname{int} }}\left( {\Br} \right) + {\tau ^{{\rm ext}}}$. For an infinite body, the internal stress can be evaluated as the convolution of the plastic strain with an elastic Green's function $G^E$, ${\tau ^{\operatorname{int} }}\left( \Br \right) = \left( {{G^E} * {\gamma ^{{\text{pl}}}}} \right)\left( \Br \right)$. Discrete plastic strain increments occur when the local stress reaches a local yield threshold $\tau^{\rm c}(\Br)$, hence the elastic domain is defined by the inequality 
\begin{equation} \label{eq:prop}
{\tau ^{{\text{th}}}}(\Br,t) = {\tau ^{\text{c}}}(\Br,t) - \left| {{\tau ^{{\text{ext}}}}(t) + \left( {{G^E} * {\gamma ^{{\text{pl}}}}(t)} \right)\left( \Br \right)} \right| \geqslant 0.
\end{equation}
The quantity $\tau^{\rm th}$ quantifies the distance of a given site from its yield stress. As long as this quantity has a positive value, the site behaves elastically. Before specifying the evolution of plastic strain which occurs once the inequality \ref{eq:prop} is violated, and the concomitant rules for assigning and evolving the local yield threshold $\tau^{\rm c}(\Br,t)$, we first need to specify the implementation of the model on a discrete lattice and explain the manner how stresses are evaluated. 

\subsection{Discretisation and stress evaluation}

The Eq.~(\ref{eq:prop}) is space-discretised on a square lattice of size $L \times L$ with periodic boundary conditions, where the edges of the square unit cell of size $d \times d$ are oriented along the $x$ and $y$ directions. To each cell we assign a single value of the local strain, the local yield stress and the local stress. Where it is not noted otherwise, distances are henceforth measured in the unit of $d$. $L$ is always an integer, and in this paper, a power of two: $L = {2^n}$.

\begin{figure}[htbp]
\begin{center}
\includegraphics[scale=0.62]{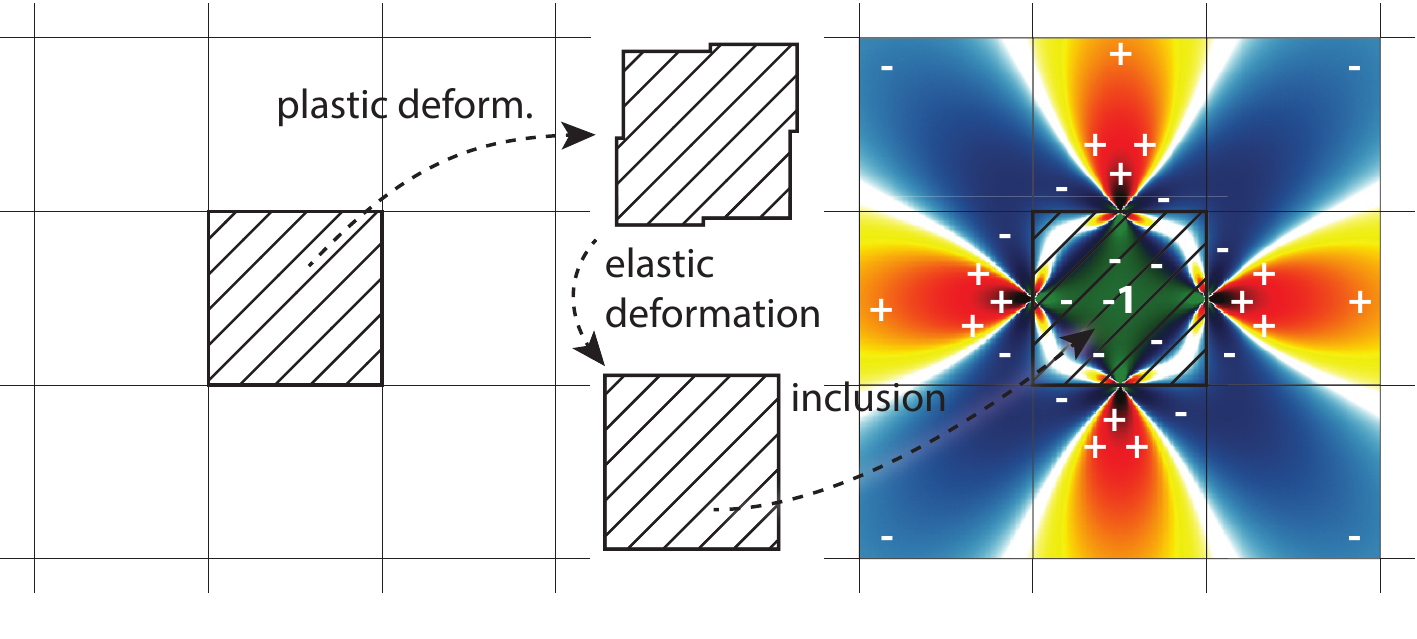}
\caption{\label{fig:stress_field_calc} In the elementary slip event, a cell is cut into 4 pieces which are displaced according to the acting shear stress and then glued back together. This cell is inserted back into the original lattice and forced elastically to fit, generating an internal stress field.}
\end{center}
\end{figure}

The stress and strain fields generated by an elementary slip event are calculated as follows (Fig. \ref{fig:stress_field_calc} illustrates the calculation.) The cell under deformation is cut along the $x$ and $y$ direction. The upper part is moved by a distance $b$ along the $x$ direction and the right side is moved by $b$ along the $y$ direction according to the sign of the shear stress acting on the cell. Then, the 4 parts are glued back together. Next, an elastic deformation is applied which transforms the cell back to its original shape so it fits its original place in the sample. The cell is placed back to its original position and the sample is elastically relaxed. The average plastic strain generated by this process in the cell is $\Delta \gamma^{\rm pl} = 2b/d$. 

\begin{figure}[htbp]
\begin{center}
\includegraphics[scale=1]{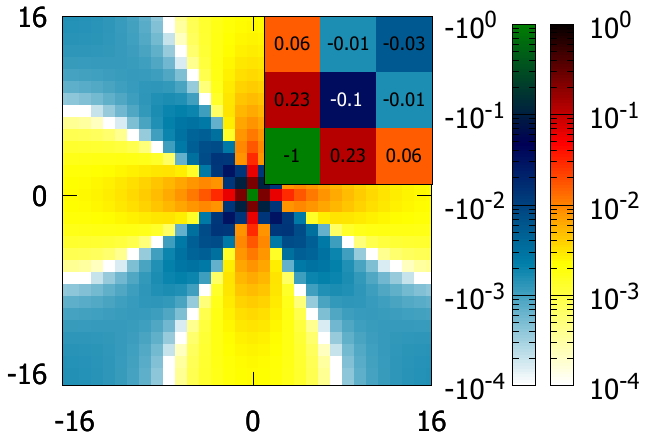}
\caption{\label{fig:stress_field} The stress field of a unit slip event located in the origin in the units of $\left| {G_{0,0}^E} \right|$, assuming periodic boundary conditions with $L = 32$. Note the symmetry in the $x$ and $y$ directions and the logarithmic color scale. In the upper right corner we show a magnification for  $k \in \left[ {0,2} \right], l\in \left[ {0,2} \right]$.}
\label{Fig:kernel}
\end{center}
\end{figure}

The process is equivalent to adding four edge dislocations\cite{Hirth1982} with the respective Burgers vectors $b{\Be_{x}}$, $b{\Be_{y}}$, $-b{\Be_{x}}$, $-b{\Be_{y}}$ at the centerpoints of the right, top, left and bottom sides of the cell. Accordingly, the stress field can be evaluated as the superposition of the stress fields of these four dislocations. Periodic boundary conditions are implemented by adding to the stress fields of the four dislocations those of their periodic images which form an infinite lattice of period $L$ (for details of the method used for evaluating the lattice sum, see \cite{Bako2006}). We evaluate stresses at the cell centerpoints, hence, the stress field induced by a elementary slip event $\Delta {\gamma ^{{\text{pl}}}}$ at the centerpoint of the active cell is $G_{0,0}^E\Delta {\gamma ^{{\text{pl}}}} =  - 2\mu \Delta {\gamma ^{{\text{pl}}}}/\left[ {\pi \left( {1 - \nu } \right)} \right]$ where $\mu$ is the shear modulus and $\nu$ is Poisson's ratio. The overall internal stress acting in an arbitrary cell $\left( {i,j} \right)$ is evaluated as 
\begin{equation}
\tau _{i,j}^{\operatorname{int}}(t) = \sum\limits_{k,l = 1}^{L} {G_{k - i,l - j}^E \gamma^{\rm pl}_{k,l}(t)},\quad
\gamma_{k,l}^{\rm pl}(t) = \sum_{t_i < t} \Delta \gamma^{\rm pl}_{k,l}(t_i). 
\end{equation}
Here $\gamma _{k,l}^{\rm pl}\left( t \right)$ is the plastic shear strain in the cell $\left( {k,l} \right)$ which is the sum of all local strain increments that have occurred in this cell up to time $t$. The kernel $G_{k,l}^E$ is the sum of the stress fields of four dislocations as detailed above, evaluated at the cell centrepoints. This method of evaluation of the internal stress field ensures that the average of the internal stress is zero, as required by stress equilibrium in an infinite body. Numerical values of the kernel $G_{k,l}^E$ in units of $\left| {G_{0,0}^E} \right|$ are shown in Fig.~\ref{Fig:kernel} for $L=32$.

The external stress is controlled by remote displacements acting on the system which impose a total (elastic and plastic) shear strain $\gamma^{\rm tot}$. Since the average of all internal stresses is by construction zero, stress equilibrium requires that 
\begin{equation}
\label{eq:tauext}
\tau^{\operatorname{ext}} = \mu \left(\gamma^{\rm tot} - \gamma^{\rm pl}  \right),\quad 
\gamma^{\rm pl} = \frac{1}{L^2} \sum\limits_{k,l = 1}^{L} \gamma^{\rm pl}_{k,l} .
\end{equation}

\subsection{Stochastic flow rule}
 
We consider a quantized, discrete plasticity model where plastic strain increases locally in discrete, finite amounts whenever the inequality, Eq.~(\ref{eq:prop}) is locally violated.  If this is the case for any site $\left( {k,l} \right)$, we increase the plastic strain  $\gamma_{k,l}^{\rm pl}$ at this site instantaneously by 
\begin{equation} \label{eq:plastic_unit}
\Delta \gamma_{k,l}^{\rm pl} = \min \left( {\Delta\gamma_0,\Delta \gamma_{k,l}^* } \right), \quad 
\Delta \gamma_{k,l}^*  = \Delta\gamma_0 \frac{\tau _{k,l}^{\operatorname{int}}  + \tau ^{{\rm ext}}}{\left| {G_{0,0}^E} \right|}. 
\end{equation}
This means that the strain in increased by a value that sets the local stress to zero if this value is less than $\Delta\gamma_0$, or otherwise by $\Delta\gamma_0$.  

Local structural disorder is taken into account in terms of random variations of the local flow threshold ${\tau ^c}(\Br)$. We assume that the system is statistically homogeneous and that the size of a cell is larger than the spatial correlation range of the microstructural disorder that gives rise to local flow stress variations. Hence, the local flow thresholds are considered as independent, identically distributed random variables ${\tau ^c}_{k,l}$ which we take to be 
Weibull distributed with exponent $\beta$ and mean value $\tau^{c}_0$. Independent values of ${\tau ^c}_{k,l}$ are initially assigned to all sites. Plasticity-induced changes in the local flow threshold are taken into account by assigning, after each local strain increment occurring at a cell, to this cell a new local flow threshold. Specifically, we draw a new value from the same distribution with average $\tau^{c}_0$ and multiply this with a strain dependent factor $F(\gamma_{k,l}^{\rm pl}) = 1 - f \gamma_{k,l}^{\rm pl}$ where $f < 0$ (called softening parameter), thus implementing a linear strain softening.

\subsection{Simulation protocol}

We non-dimensionalize the model by measuring all stresses in units of the mean flow threshold $\tau^c_0$, all strains in units of $\tau^c_0/\mu$ (elastic strain needed to reach the mean flow threshold, divided by the shear modulus), and spatial coordinates in units of the cell size $d$. The model behaviour is then, in addition to the Weibull parameter $\beta$, controlled by a single numerical parameter $I = 2 \mu \Delta\gamma_0/[\pi(1-\nu)\tau^c_0]$ (henceforth: `coupling constant') which controls the magnitude of the internal-stress re-distribution after a deformation event relative to the average flow stress. In the following we make the simplifying assumption that $\nu = 0.353$ in which case $I=\mu \Delta \gamma_0/\tau^c_0$ equals the scaled local strain increment. The local stress reduction at the site of a unit deformation event is then $I$ and the external stress reduction associated with the same event is $I/L^2$. 

Simulations are performed as follows: We assign initial flow thresholds to all sites according to the prescribed Weibull distribution with exponent $\beta$ and mean 1. We then determine the site with the lowest threshold and increase the total strain $\gamma^{\rm tot}$ such that the concomitant stress increase as given by Eq.~(\ref{eq:tauext}) exactly matches the threshold, triggering the first deformation event. After the event, which is supposed to occur instantaneously, we re-compute all stresses while keeping $\gamma^{\rm tot}$ fixed, evaluate the local threshold stresses $\tau^{\rm th}_{k,l}$ for all sites, and check whether there are additional sites which become unstable ($\tau^{\rm th}_{k,l} < 0$). If yes we increase, still at fixed $\gamma^{\rm tot}$, the strain at the unstable site with the lowest value of $\tau^{\rm th}_{k,l}$, thus implementing an extremal dynamics. We repeat this until there are no more unstable sites (the avalanche has terminated). The plastic strain and the stress at this point are evaluated from Eq.~(\ref{eq:tauext}). We then determine again the site with the smallest threshold, increase $\gamma^{\rm tot}$ such that the concomitant stress increase as given by Eq.~(\ref{eq:tauext}) makes this site unstable and triggers the next avalanche. We repeat this cycle of avalanche triggerings until the local strain of at least one site reaches the value $\gamma_{k,l}^{\rm pl} = 1/f$ such that the strength of this site becomes zero. This is tantamount to the nucleation of a microcrack which we take as a signature of impending system failure. The concomitant average plastic strain defines the system failure strain $\gamma^{\rm pl}_{\rm f}$.

\section{Results}
Simulations were performed for Weibull shape parameters $\beta = 1, 2, 4, \text{ and } 8$, coupling constants $I = 0.125,0.25,0.5 \text{, and } 1$, and for system sizes $L = 32, 64, 128, 256 \text{ and } 512$. In each case ensembles of 512 simulations with statistically independent initial conditions were performed. The softening parameter $f$ was kept fixed at $f = 1/16$. 

\subsection{Stress-strain curves}
\begin{figure}[htbp]
\begin{center}
\includegraphics[scale=0.5, angle=0]{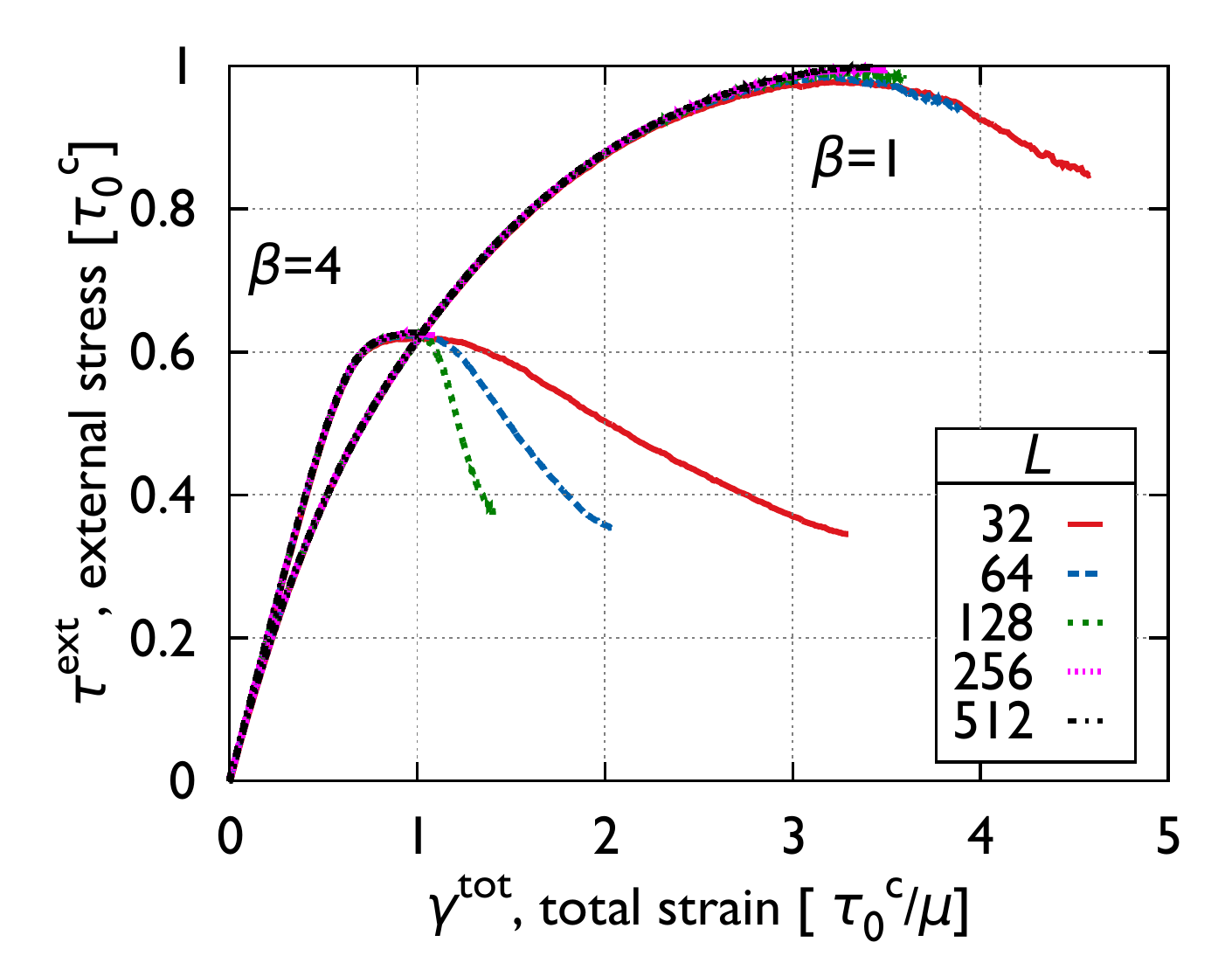}
\caption{\label{fig:ssc} Stress-total strain curves for two different yield-stress distributions (Weibull exponents $\beta= 1$ and $\beta=4$) and different system sizes; other parameters $I=1; f=1/16$.}
\end{center}
\end{figure}

Average stress-strain curves were obtained by averaging the external stress at a given deformation over the simulations as shown in  Fig. \ref{fig:ssc}. 

The curves exhibit three different regimes: An initial quasi-elastic loading regime is followed by a transition to a plastic deformation regime where the stress increases with strain (hardening regime). The elastic and hardening regimes are system size independent. The hardening regime is followed by a transition to a softening part where the stress decreases with macroscopic strain. The simulations are terminated once microcrack nucleation occurs as indicated by a complete loss of strength at one or more sites. The corresponding failure strains are much below the expectation $\gamma_f = 16$ for a homogeneous system, indicating a significant degree of deformation localization. We also observe that the softening regime is system size dependent: The stress decrease occurs more rapidly and failure occurs at lower strains in larger systems. Such system size dependence again indicates some kind of deformation localization. We therefore proceed to investigate the strain patterns that emerge in the different deformation stages. 

\subsection{Patterns in the strain maps}

\begin{figure}[htbp]
\begin{center}
\includegraphics[scale=0.23, angle=0]{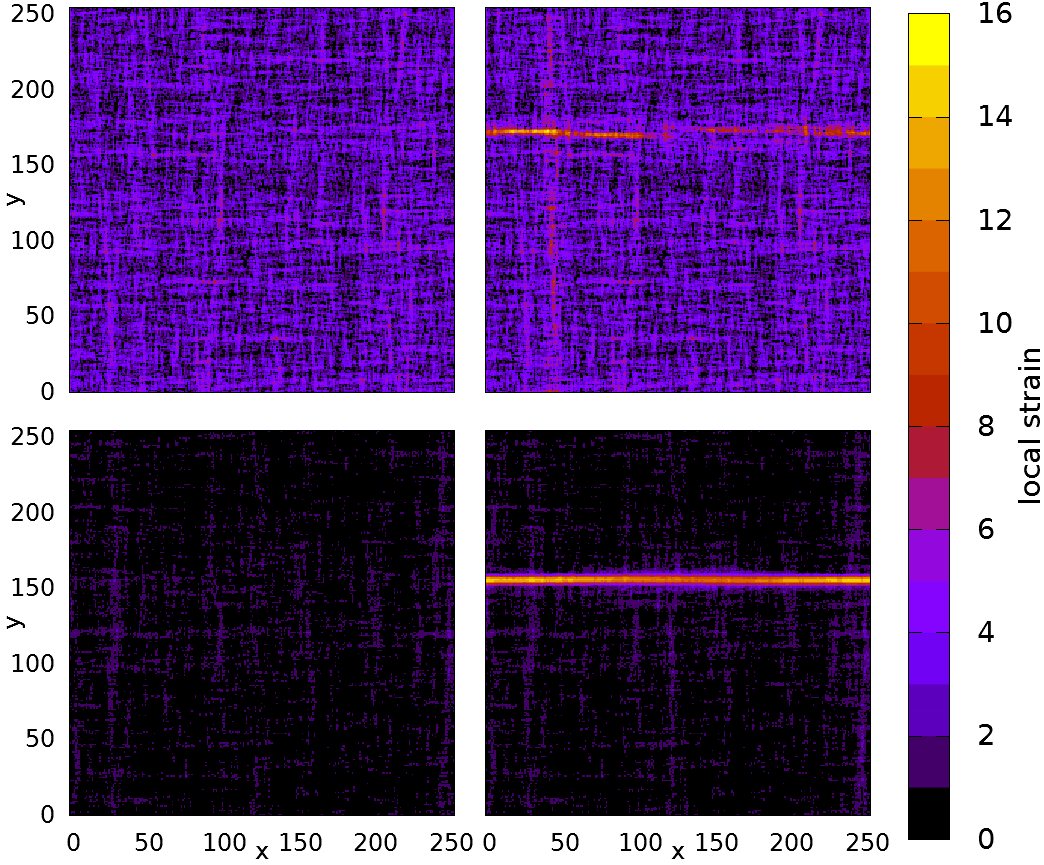}
\caption{\label{fig:pattern} Strain patters at the highest external stress just before the onset of softening (left) and at the end of the simulation (right); 
$\beta=1$ (top) and $\beta=4$ (bottom); other parameters: $I = 1, f=1/16, L=256$.}
\end{center}
\end{figure}

Figure \ref{fig:pattern} illustrates the changes in the strain patterns that occur during the softening regime. At the peak stress before the onset of softening, deformation is macroscopically homogeneous but exhibits mesoscale patterns in the form of numerous diffuse shear bands which follow the planes of maximum shear stress, here aligned with the $x$ and $y$ directions. These patterns are more pronounced with increasing degree of disorder. Note that the peak stress is reached later in the more disordered sample (top left graph in Fig.~\ref{fig:pattern}), hence the overall strain is bigger. During the softening regime we observe a qualitative change in the patterns as most of the additional strain accruing during the softening regime is localized in a single shear band which also contains the location where microcrack nucleation takes place. This shear band is sharper and more pronounced in the sample with less disorder (bottom right graph in Fig.~\ref{fig:pattern}). 

The formation of a localized shear band is in line with the ideas of classical continuum mechanics which predicts localization to occur, in a system without boundary constraints and under pure shear loading, at the transition from strain hardening to strain softening regimes. To better characterize this behavior we now seek to define a quantitative measure for the strain localization process. 

\subsection{Deformation localisation}

In order to quantify strain localisation we investigate the spatial distribution of the incremental strain. 
We divide the average stress-strain curve into $n=50$ intervals, the $k\text{th}$ interval is defined by
${\gamma ^{{\text{pl}}}} \in \left[ {{\gamma ^{{\text{pl}},k}},{\gamma ^{{\text{pl}},k + 1}}} \right),\quad {\gamma ^{{\text{pl}},k}} = k\left\langle {\gamma _{\text{f}}^{{\text{pl}}}} \right\rangle /n$. The plastic strain increment occurring at the site $(i,j)$ during strain interval $k$ is denoted as $\Delta \gamma _{i,j}^{{\rm pl},k}$. 

We now use that a shear band has a planar shape. For any given plane $\cal P$ we can define a scalar measure of distance which characterizes the 
distribution of the incremental strain with respect to the plane. To this end we denote the distance between site $(i,j)$ and the plane $\cal P$ 
as $d_{i,j}^{\cal P,\perp}$. (Because of the periodic boundary conditions used, we evaluate $d_{i,j}^{\cal P,\perp}$ as the minimum distance between the 
site $(i,j)$ and any of the periodic images of $\cal P$). We now define
\begin{equation}
d_{\gamma,k}^{\cal P} = \frac{\sum_{ij} \Delta \gamma _{i,j}^{{\rm pl},k} \cdot d_{i,j}^{\cal P \perp}}{\sum_{ij} \Delta \gamma _{i,j}^{{\rm pl},k}}.
\end{equation}
For a completely homogeneous distribution of the plastic strain increment, we have $d_{\gamma,k}^{\cal P} = L/4$ for all planes $\cal P$. For a heterogeneous distribution we identify the plane for which $d_{\gamma,k}^{\cal P}$ has the smallest value and define a localization parameter $\eta$ as 
\begin{equation}
\eta_k = 1- \frac{4 d_{\rm min,k}}{L} \quad,\quad d_{\rm min,k} = \min_{\cal P} d_{\gamma,k}^{\cal P} 
\end{equation}
This parameter takes the value $\eta_k = 0$ for a statistically homogeneous distribution of the plastic strain increment, and the value 
$\eta_k = 1$ if the incremental strain is completely localized on a single plane. 

\begin{figure}[htbp]
\begin{center}
\includegraphics[scale=0.5, angle=0]{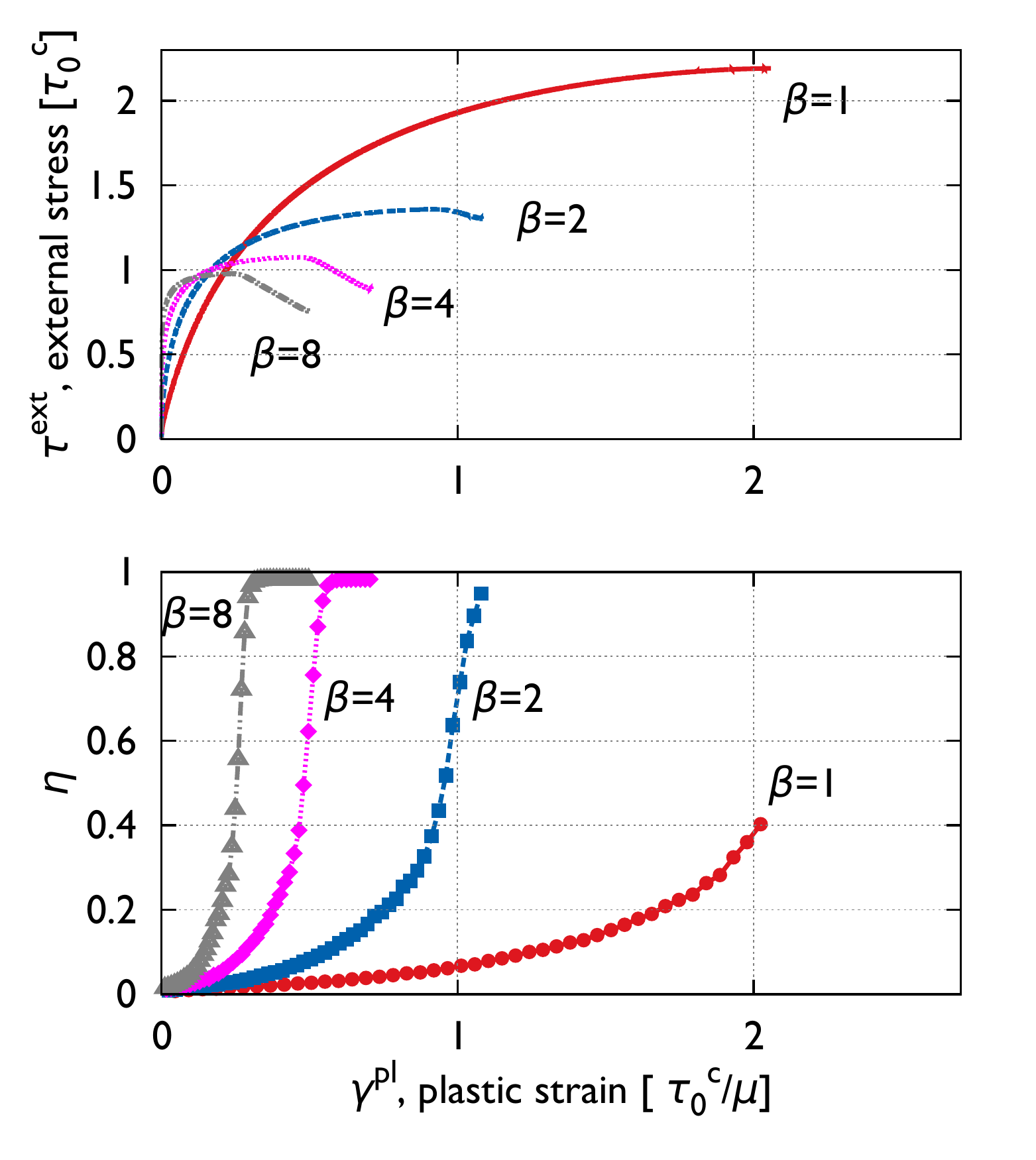}
\includegraphics[scale=0.5, angle=0]{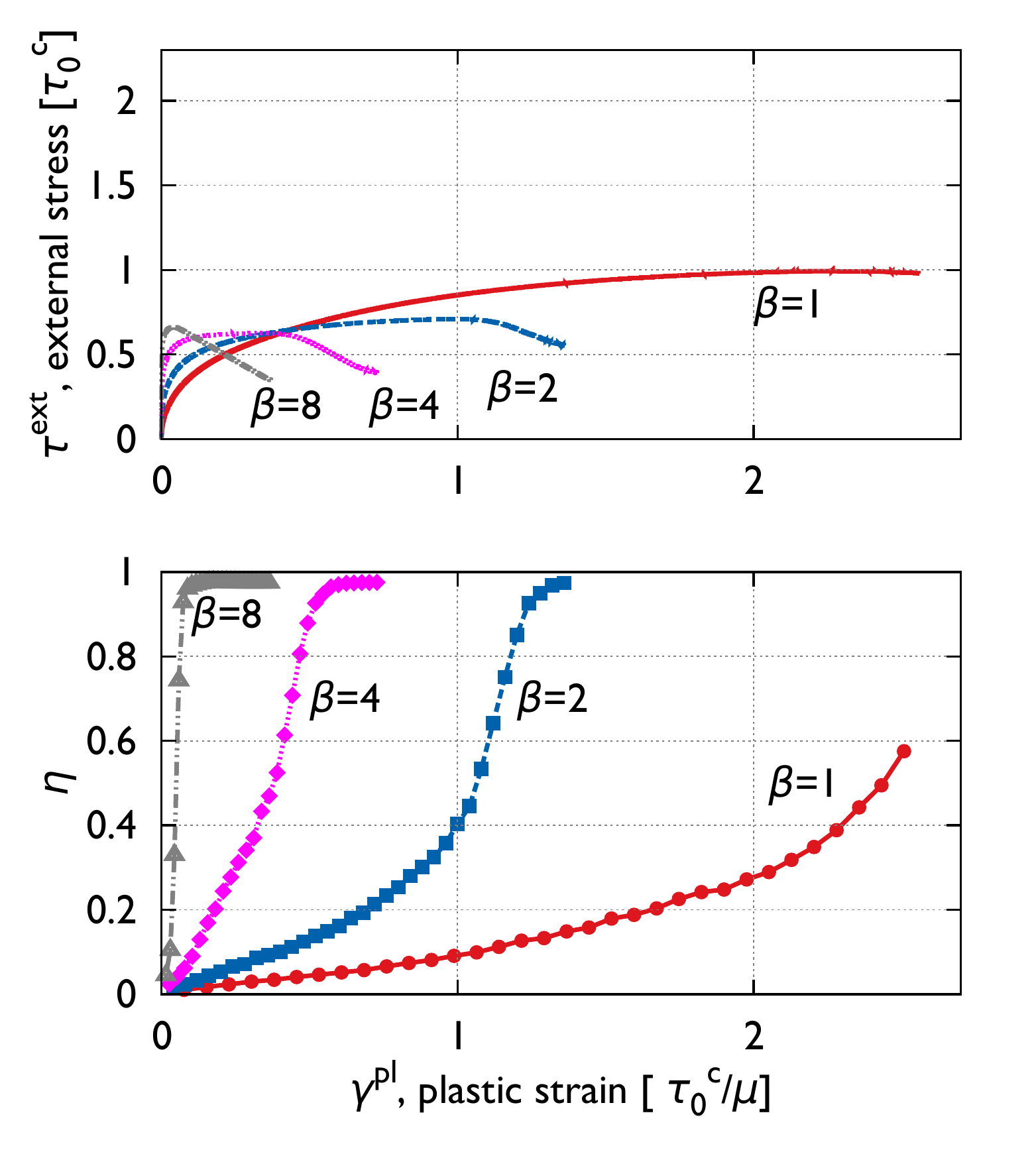}
\caption{\label{fig:radius} Stress strain curves and strain evolution of the localization parameter $\eta$
for different degrees of disorder (Weibull parameter $\beta \in [8,4,2,1]$. 
The upper figure corresponds to $I=0.125$, the lower to $I=1$.}
\end{center}
\end{figure}

It can be seen in Fig~\ref{fig:radius} that in all simulations the localization parameter $\eta$ starts at $\eta = 0$ and then gradually increases during the hardening regime. Immediately after the peak stress is reached and the system enters the macroscopic softening regime, $\eta$ increases rapidly towards $\eta = 1$, indicating the localization of deformation in a single shear band. The comparison of the strain evolution of $\eta$ and $\tau^{\rm ext}$ clearly demonstrates the correlation. It is equally evident that an increasing degree of disorder (decrease of the Weibull exponent from $\beta = 8$ to $\beta = 1$), even though it leads to an earlier onset of plastic flow, extends the hardening regime to larger strains and delays the onset of deformation localization. The role of the coupling constant $I$, which reflects the magnitude of the local strain increment, is more ambiguous: for small disorder, large values of $I$ seen to promote localization whereas for large disorder, the opposite is the case.

\begin{figure}[htbp]
\begin{center}
\includegraphics[scale=0.5, angle=0]{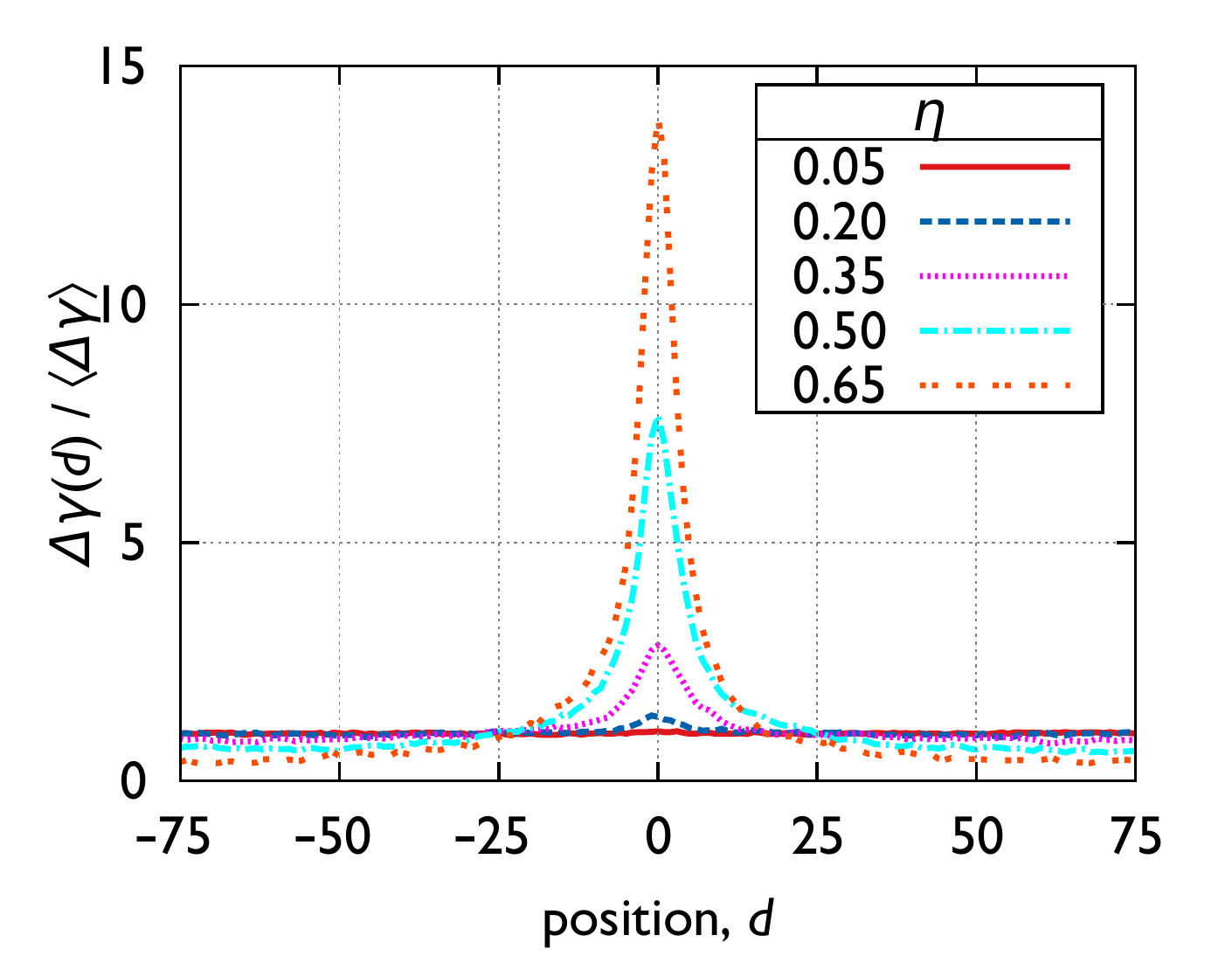}
\includegraphics[scale=0.5, angle=0]{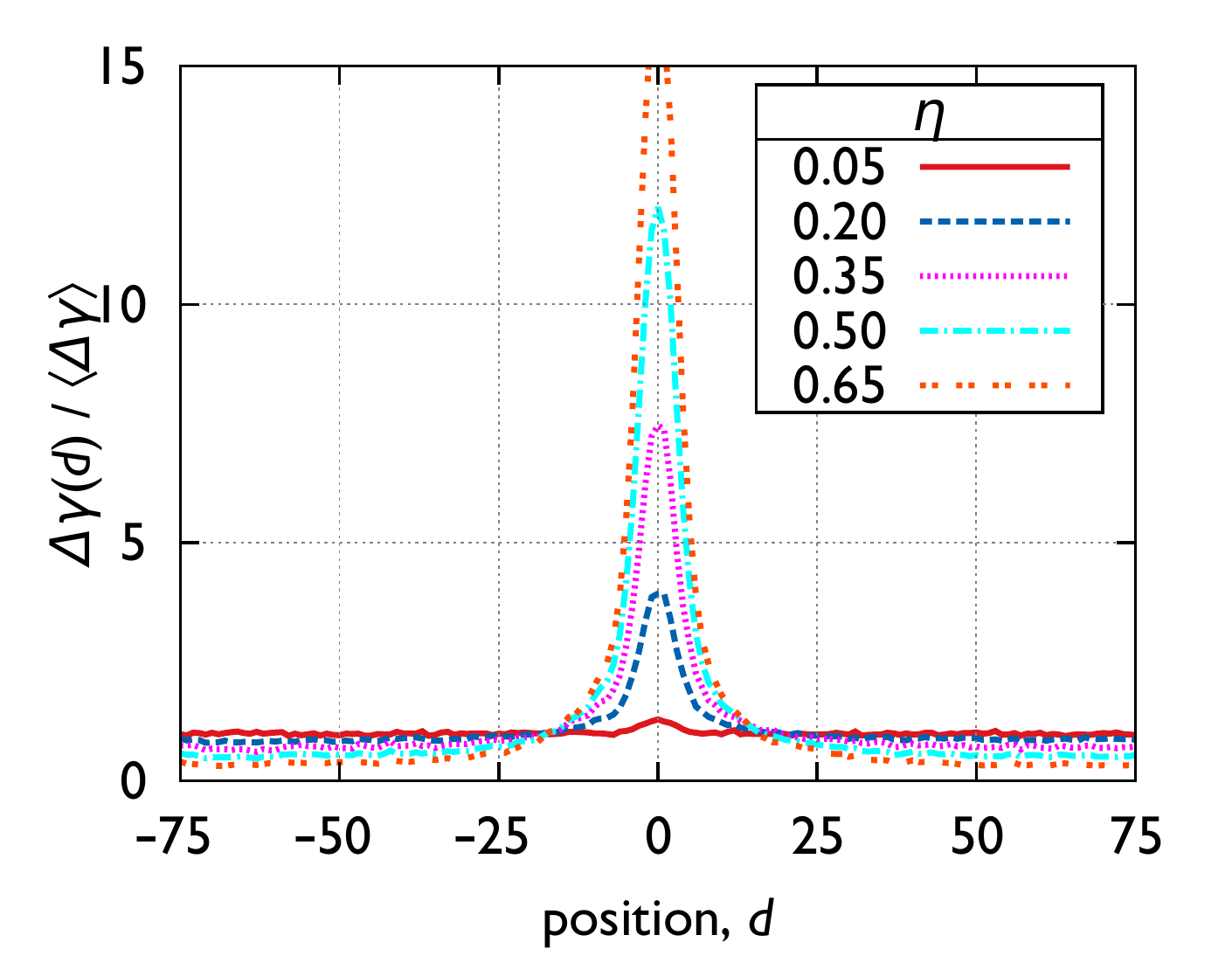}
\caption{\label{fig:shearband} Evolution of the distribution of strain around the final failure plane for Weibull parameters $\beta = 1$
(top) and $\beta=8$ (bottom).}
\end{center}
\end{figure}

We now look at the distribution of incremental strain around the {\em final} failure plane. This is shown 
in Fig.~\ref{fig:shearband} which depicts shear band profiles for large disorder ($\beta = 1$) and for small disorder ($\beta = 8$), recorded for different values of the localization parameter $\eta$. The width of the shear band is almost the 
same in both cases, however, localization of deformation around the final failure plane happens later in case of large disorder. This looks strange at first glance, given that the curves compare situations with equal value of $\eta$, however, the reason is simple: In case of large 
disorder, deformation first localizes in a transient manner (i.e., on slip planes that are in general {\em not} close to the final failure plane) and localization on the final failure plane happens after extensive deformation activity has occurred elsewhere. In case of small disorder, by constrast, deformation localizes on the final failure plane almost from the onset.

\subsection{Mean strain to failure}

The beneficial effect of disorder on ductility is also borne out when we consider the mean strain at failure. 
\begin{figure}[htbp]
\begin{center}
\includegraphics[scale=0.5, angle=0]{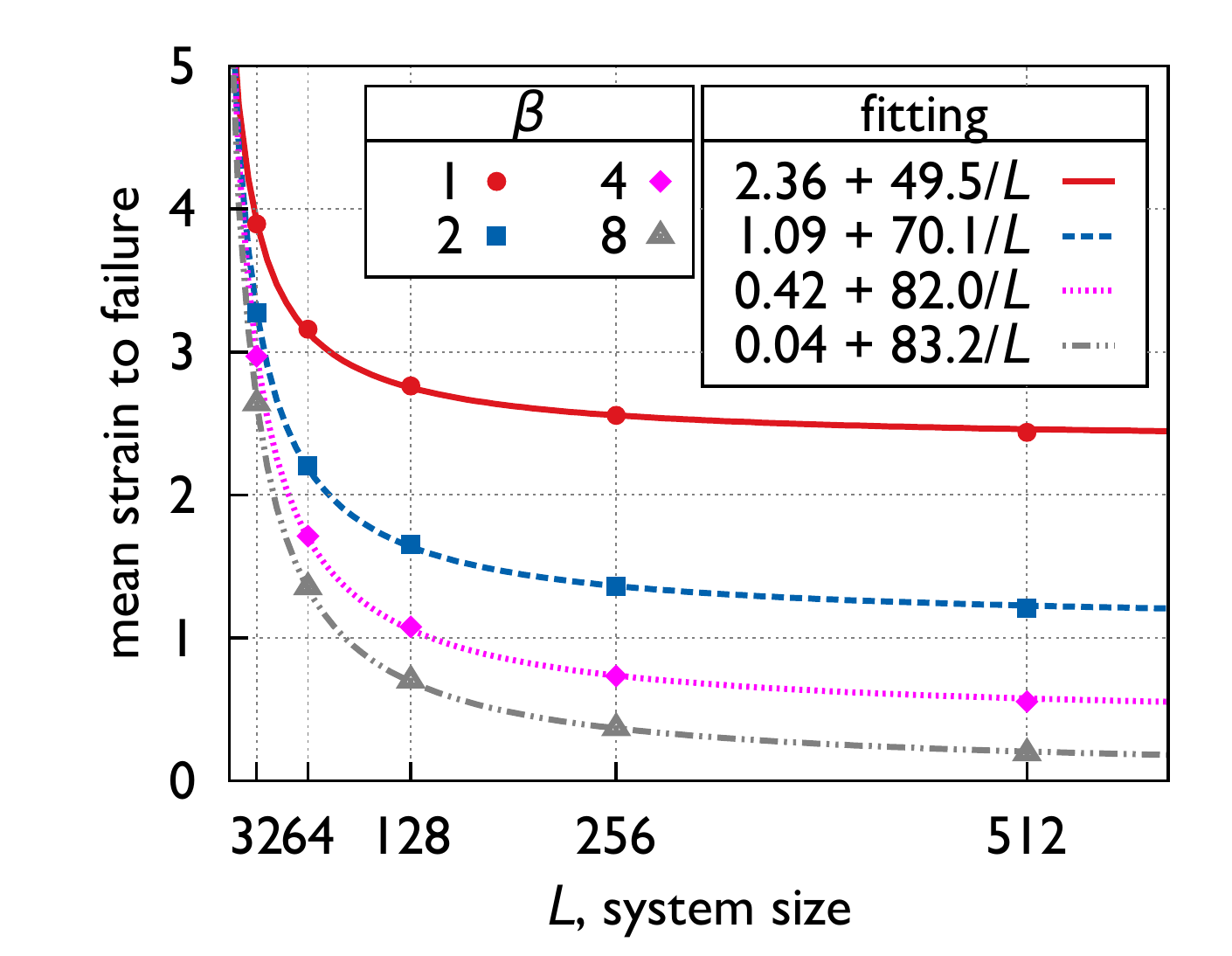}
\caption{Mean strain at failure as a function of system size, for different Weibull parameters; dashed lines: 
fit curves $\gamma_f = c_1 + c_2/L$}.  
\label{fig:strain_to_failure_size}
\end{center}
\end{figure}
This strain is system size dependent and decreases with increasing system size. This dependency can be rationalized by making the simplifying assumption that the system deforms homogeneously during the hardening regime, accumulating a homogeneous strain $\gamma_h$, whereas all strain accruing during the subsequent softening regime is localized in a slip band of finite width $d_{\beta}$. Failure occurs once the plastic strain in the band reaches the value $\gamma_{\rm f}^{\rm loc}$. Then the mean strain at failure is $\gamma_f = \gamma_h + (\gamma_{\rm f}^{\rm loc}  - \gamma_h)(d/L)$. We can thus fit the system size dependence as $\gamma_f = c_1 + c_2/L$ where the parameter $c_1$ defines the homogeneous strain $\gamma_h$ which is also the failure strain in the infinite system limit. This strain is plotted in Fig.~\ref{fig:strain_to_failure_shape} as a function of the Weibull exponent $\beta$. Again we see that larger microstructural disorder leads to an increase in ductility of the strain softening material. For
large systems $(L \to \infty)$ the increase is quite dramatic - between Weibull exponent $\beta =8$, corresponding to a coefficient of variation of 0.148, and Weibull exponent $\beta =1$ (coefficient of variation 1), the strain to failure increases in this limit by a factor of about 60. 

\begin{figure}[htbp]
\begin{center}
\includegraphics[scale=0.5, angle=0]{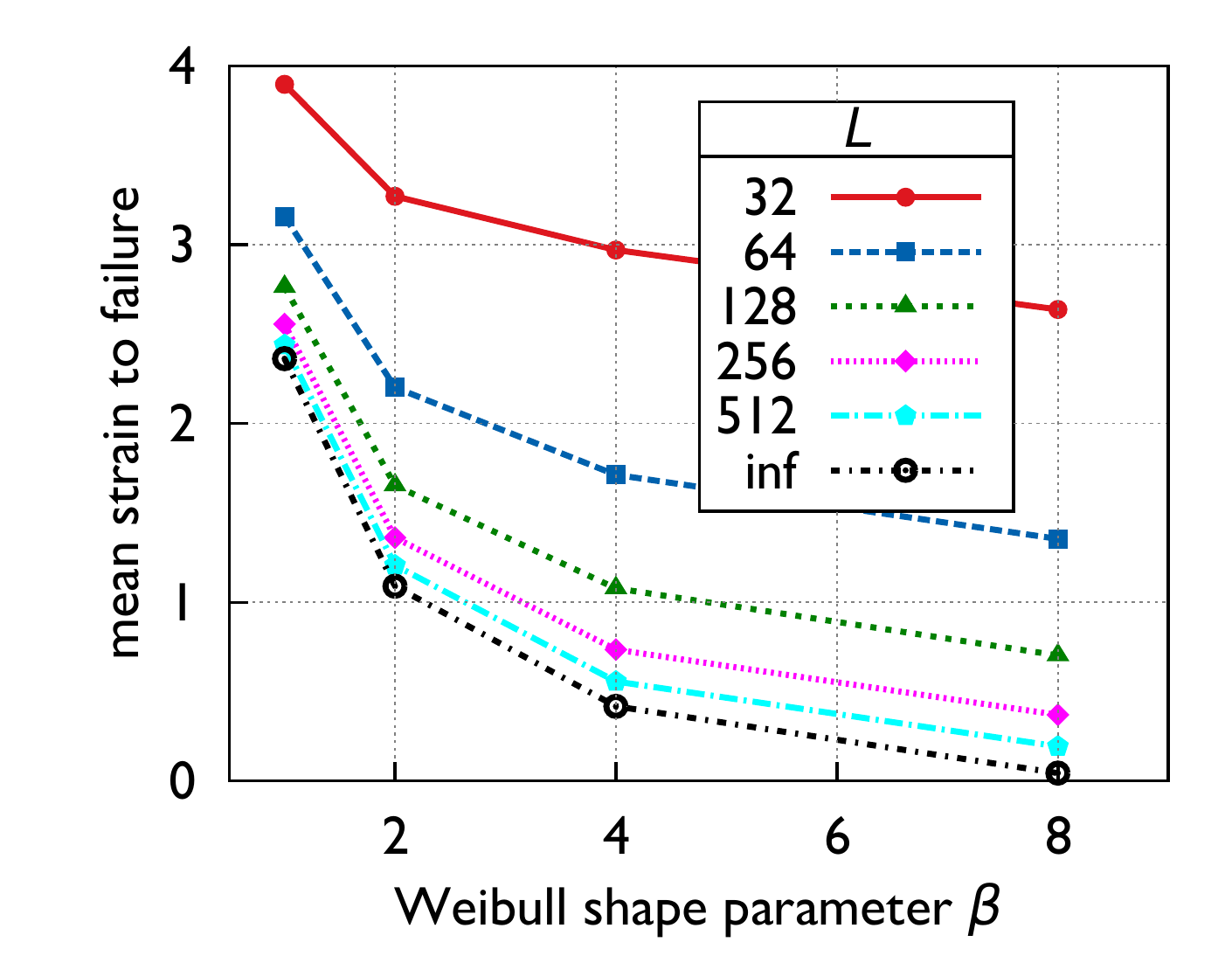}
\caption{Mean strain at failure as a function of the Weibull shape parameter $\beta$, for different system sizes. The value for infinitely large system size is obtained from the parameter $c_1$ of the fit curves in Figure \ref{fig:strain_to_failure_size}.}
\label{fig:strain_to_failure_shape}
\end{center}
\end{figure}

\subsection{A local criterion for shear band growth}

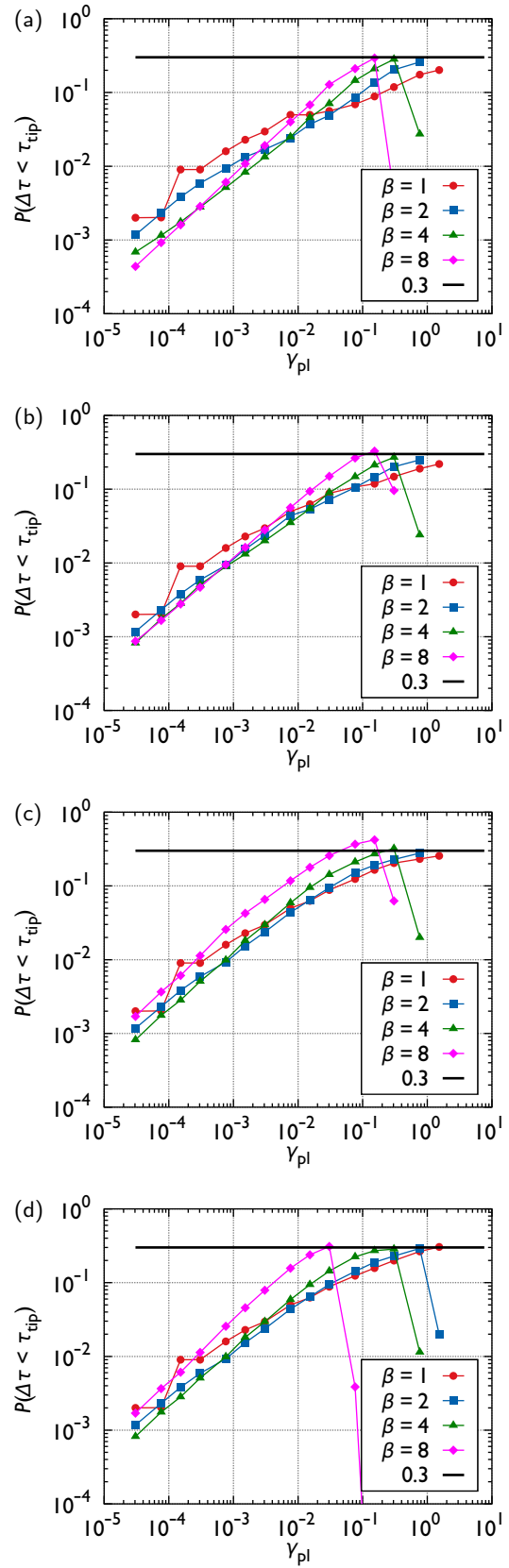
\begin{figure}[htbp]
\begin{center}
\includegraphics[scale=0.5, angle=0]{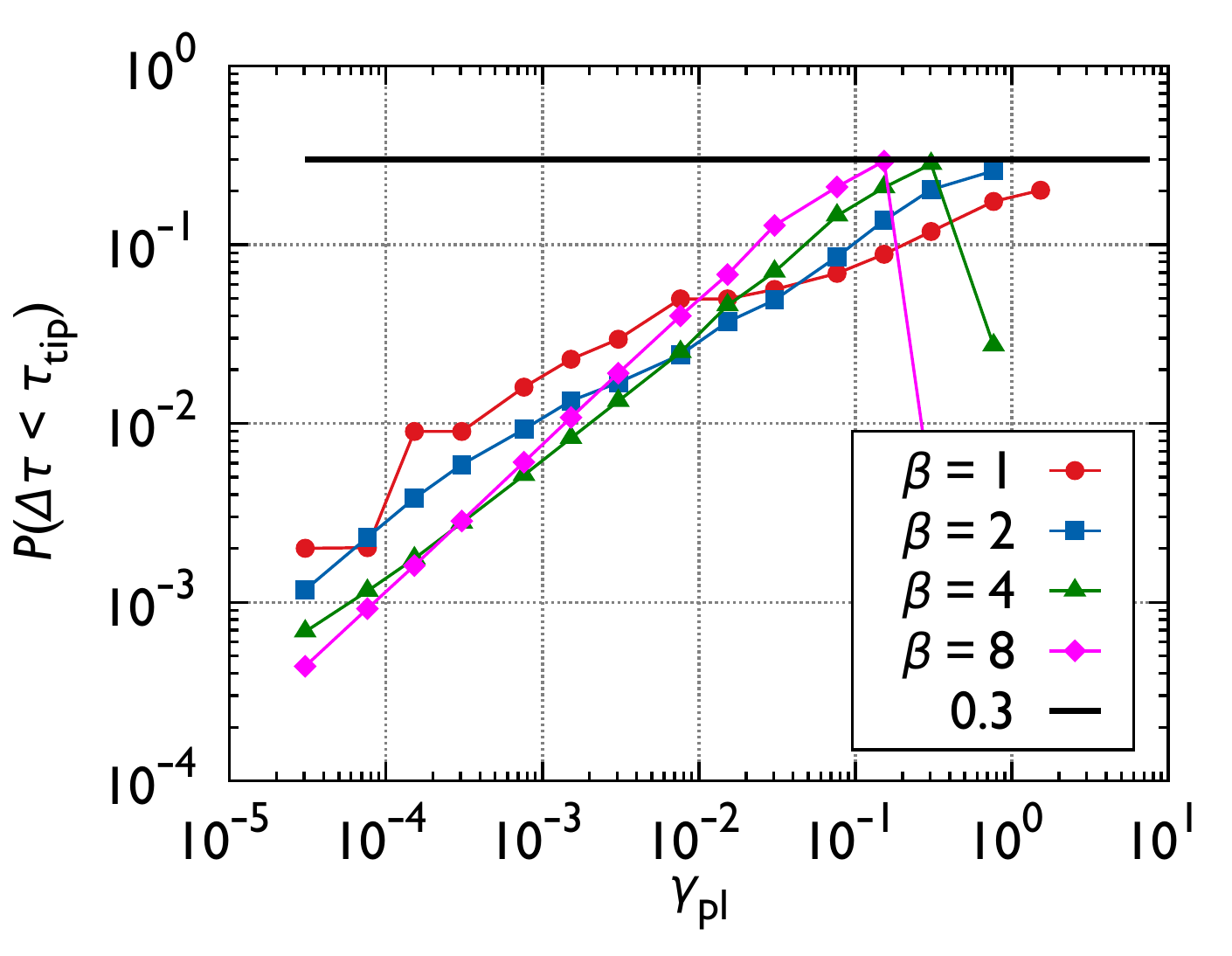}
\begin{picture}(0,0)
 \put(-205,144){\textsf{(a)}}
\end{picture}
\includegraphics[scale=0.5, angle=0]{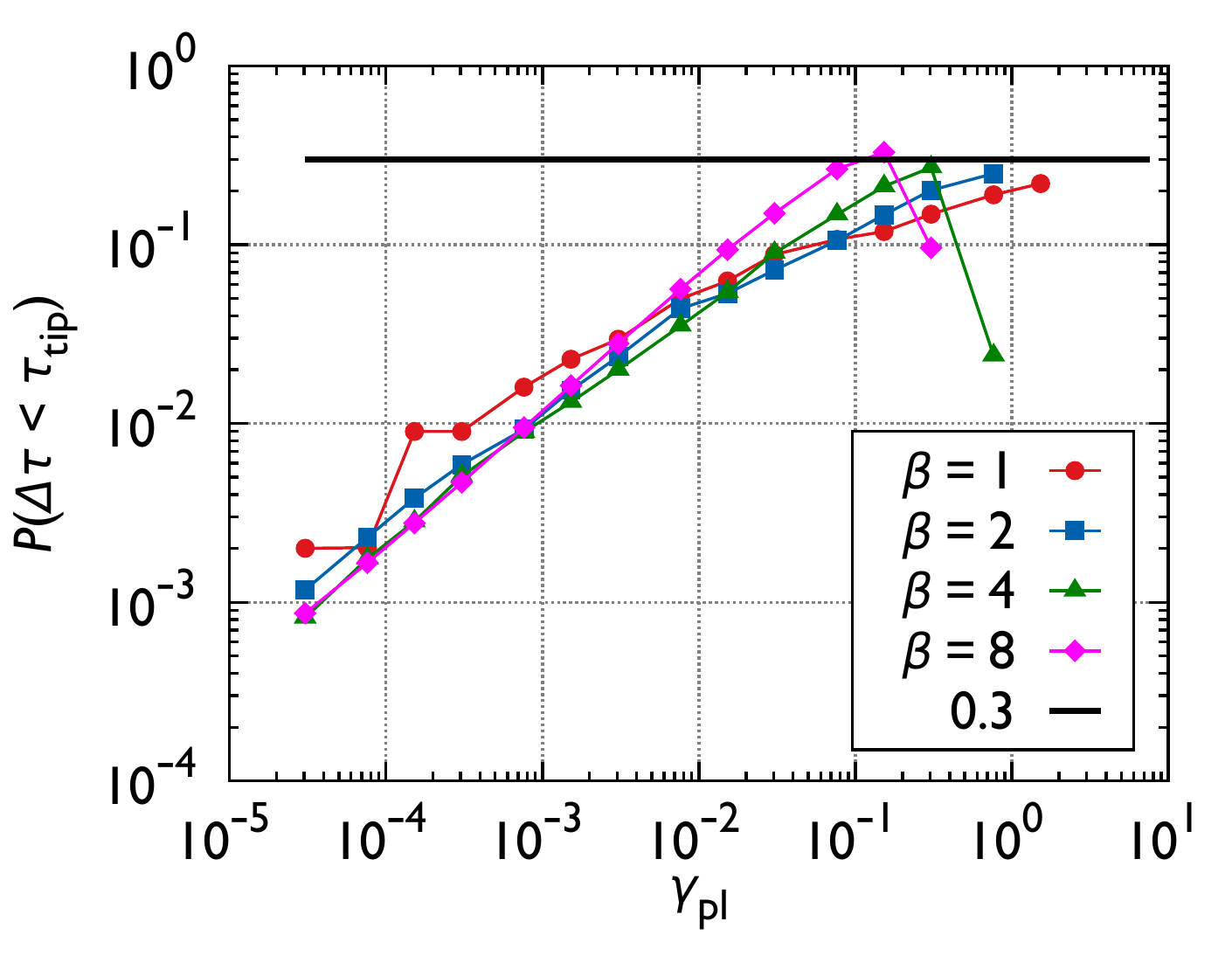}
\begin{picture}(0,0)
 \put(-205,144){\textsf{(b)}}
\end{picture}
\includegraphics[scale=0.5, angle=0]{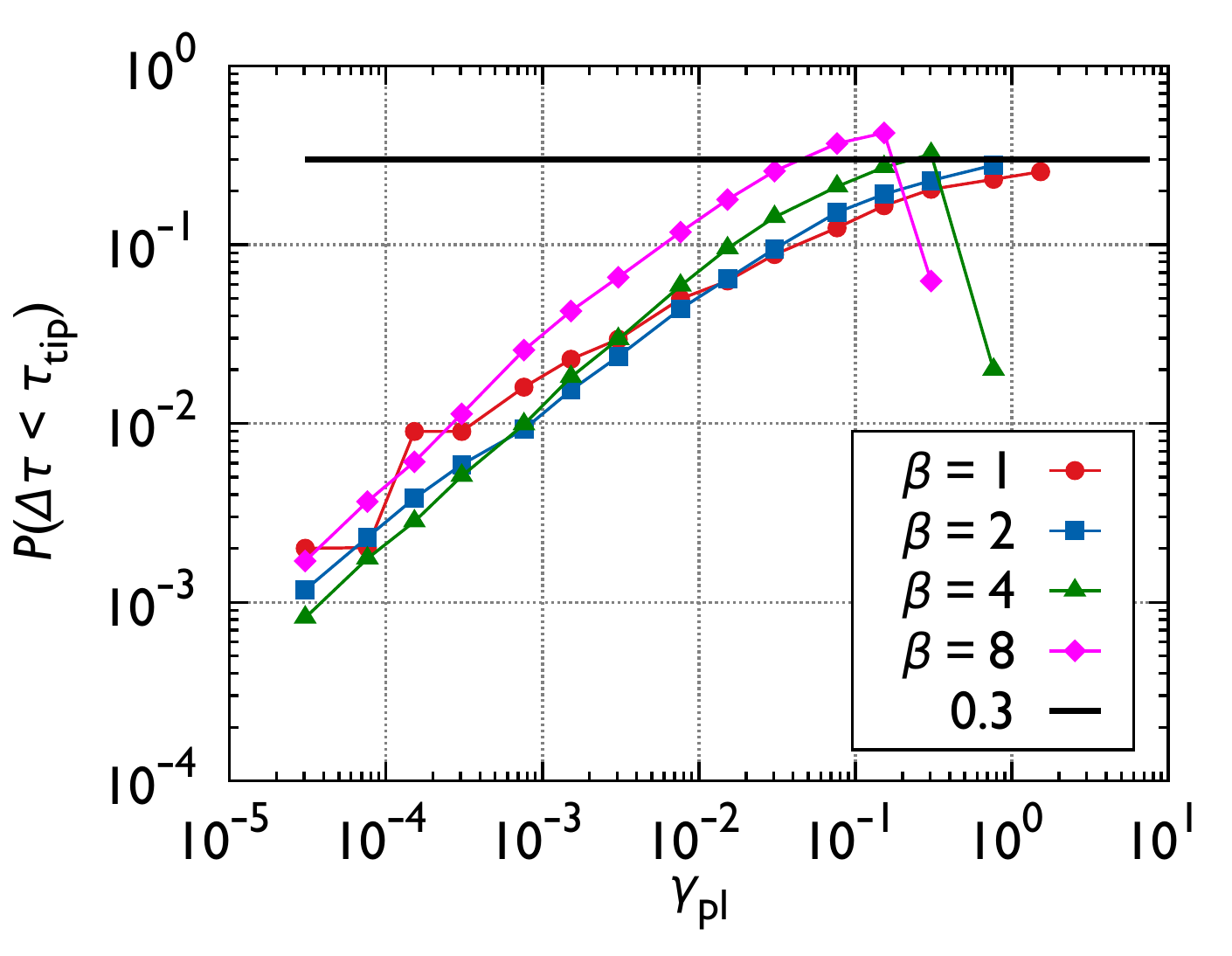}
\begin{picture}(0,0)
 \put(-205,144){\textsf{(c)}}
\end{picture}
\includegraphics[scale=0.5, angle=0]{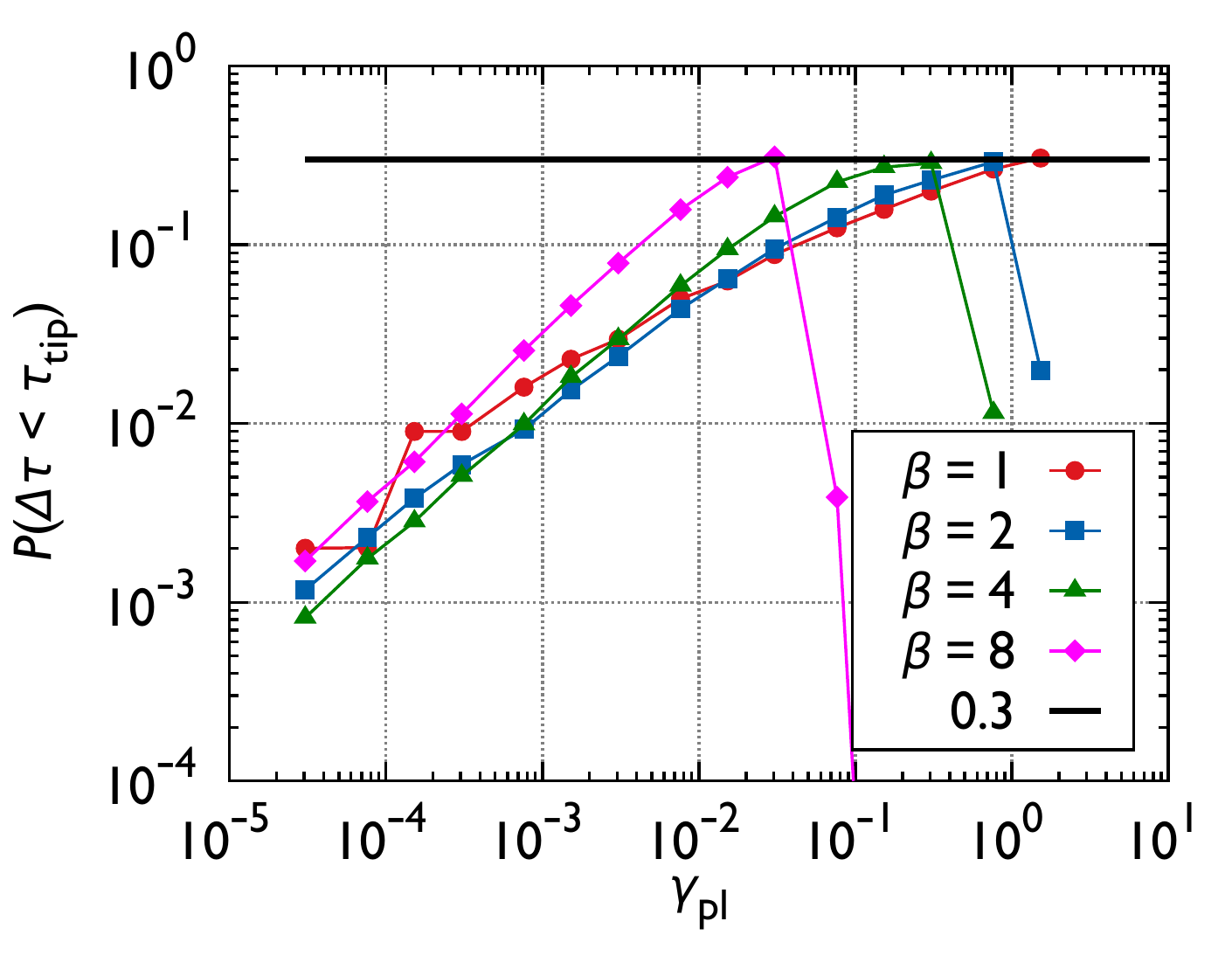}
\begin{picture}(0,0)
 \put(-205,144){\textsf{(d)}}
\end{picture}
\caption{\label{fig:trigger} Evolution of the triggering probability $P(\Delta \tau < \tau_{\rm tip})$ at the tip of an incipient shear band. 
(a) $I=0.125$, (b) $I=0.25$, (c) $I=0.5$ (d) $I=1$.}
\end{center}
\end{figure}

Failure of a macroscopic system occurs once the first macroscopic (system spanning) shear band forms and deformation localizes there. However, embryonic shear bands are present already before the onset of softening (Fig.~\ref{fig:pattern}). At this point we ask whether we can establish a criterion which allows us to better understand the conditions for the emergence of a macroscopic shear band. To this end we assume a pre-existing shear band of some extension and investigate its growth. The stress concentration at the tip of a straight band with a width of one cell can be estimated as
\begin{equation}
\tau_{\rm tip} = \sum_{k=1}^{\infty} G^E_{k,0} \approx 0.385 G^E_{0,0}
\end{equation}
The band expands if this stress concentration triggers a deformation event at one of the sites ahead of either tip. We denote the stress needed to activate a site with $\Delta\tau$ (we might also call this the residual strength of the site), and the probability that a randomly chosen site is activated by a stress increment $\Delta \tau < \tau^*$ as $P(\Delta \tau < \tau^*)$. We now investigate the evolution of the probability $P(\Delta \tau < \tau_{\rm tip})$ (i.e, the probability that an advance of a band triggers another advance straight ahead) as a function of strain. Figure \ref{fig:trigger} indicates that this probability increases continually with increasing strain until it reaches a level of about $P(\Delta \tau < \tau_{\rm tip}) \approx 0.3$ which does not depend on the model parameters (disorder  parameter $\beta$, coupling constant $I$). At this critical value, $P(\Delta \tau < \tau_{\rm tip})$ suddenly drops. This drop coincides with the formation of a system spanning shear band where deformation localizes: the associated stress drop reduces the value of $P(\Delta \tau < \tau_{\rm tip})$ everywhere except in the b
and itself where the local strain softening maintains it at the critical level for sustained shear band operation. 

Despite significant variations in the distributions $P(\Delta \tau < \tau^*)$ which depend strongly on the disorder parameter $\beta$, we observe that critical value of $P(\Delta \tau < \tau_{\rm tip})$ is quite universal. This leads us to the conclusion that shear band growth is 
driven by processes occurring at the tip of an incipient shear band,  and that it depends on the interaction of the shear band tip with the local distribution of the residual strength whether  or not shear band growth can occur. Indeed, if we account for the fact that shear bands may grow at both tips, and that growth may not necessarily be constrained to expansion straight ahead but may occur via sideways deflection (thus, at each tip there may be three sites available for continuing growth), we can estimate that a critical probability of the order of 1/3 for triggering a site at the crack tip may be sufficient for sustained growth of a shear band.   

Thus our analysis leads us to envisaging shear band formation as essentially a two-stage process: In a first stage, local yielding and the concomitant stress re-distribution lead to a system-wide re-shuffling of the residual strength distribution in such a manner  that the triggering probability $P(\Delta \tau < \tau_{\rm tip})$ at the tip of an incipient shear band gradually increases. During this stage, deformation is macroscopically homogeneous, even though shear band nuclei are continuously forming and getting again inactivated. The duration of this latency stage depends on the degree of disorder and increases with increasing disorder parameter $\beta$. As soon as the triggering 
probability reaches a critical value $P(\Delta \tau < \tau_{\rm tip}) \approx 0.3$, a transition to a second stage occurs where the flow process is governed by the rapid formation of a system-spanning shear band where deformation localizes. In large systems, this leads to near-instantaneous failure. 

It may be noted in passing that the residual strength probability distribution $P(\Delta \tau < \tau^*)$, notably its behavior near the edge of stability, $\Delta \tau \to 0$, has recently been conjectured by Wyart and co-workers \cite{Lin2014,Lin2015} to play a crucial role in the dynamics of similar models as the present one but {\em without} softening. In these works it is argued that the spatial organization of slip in shear bands is largely irrelevant during the approach to failure, and that the non-local elastic kernel can be approximated by a random stress re-distribution in a mean-field model which by construction destroys any spatial correlations. Our observations, by contrast, point to the importance of correlated shear band growth in controlling system stability as soon as some degree of softening is introduced into the model.

\section{Discussion and Conclusions}

We studied the deformation and failure behavior of microstructurally disordered model materials which exhibit irreversible strain softening and therefore fail by shear band formation. Contrary to the intuitive idea that increased microstructural heterogeneity may facilitate shear band nucleation and therefore have a negative impact on deformability, we find a strong positive effect of increased heterogenity and randomness on the deformation properties. Increased microstructural heterogeneity indeed leads to an earlier onset of deformation in the form of diffuse shear bands -- an effect that is easily understood within the classical paradigm of weakest-link statistics (for a discussion in the plasticity context see e.g. \cite{Ispanovity2013}. However, the same heterogeneity prevents the spreading of shear bands and their coalescence into a system spanning macro-shear-band. The earlier onset of deformation is matched by an extended hardening regime, associated with the elimination of weak regions from the microstructure. This hardening can be envisaged as a survival-bias-hardening (easily deformable configurations are eliminated, stronger configurations survive) and becomes more pronounced with increasing scatter in local strength. Only after the survival-bias-hardening is exhausted, structural softening takes over and promotes macroscopic deformation localization. In line with classical concepts of continuum mechanics of homogeneous materials, the onset of macroscopic localization neatly coincides with the peak stress where the system enters the softening regime of the stress strain curve. 

Our findings indicate that, in microstructurally disordered materials where ductility is limited by shear band formation, it may be a good idea to {\em increase} the degree of microstructural heterogeneity - an increase which results both in an increase in strength and in a very significant increase in ductility. In the context of metallic glasses, our findings match well with ideas to increase the ductility of metallic glasses by introducing a second interface phase \cite{Adibi2013} or by embedding nano-crystallites into a glassy matrix \cite{Das2005} -- ideas which are tantamount to increasing the scatter of local deformation properties within a generally disordered microstructure. 

Of course, it is a well established idea that strong-yet-ductile materials can be engineered by combining weak-but-ductile and strong-but-brittle components into heterogeneous composite microstructures. However, this is not what we are studying in the present work: As manifest from the constitutive relation of our model material, volume elements of different strength are assumed to fail at the same local strain. If we investigate the evolution of the final, macroscopic shear band, then we can see little difference between weakly disordered and strongly disordered microstructures (Fig.~\ref{fig:shearband}). Nevertheless, the overall deformation behavior is radically different in both cases, because the disorder extends the hardening regime and prevents local shear band nuclei from coalescing into a macroscopic shear band. To understand this behavior it is necessary to go beyond the investigation of averaged, effective materials properties and move towards an understanding of the manner how fluctuations emerge and extend across scales. This viewpoint is corroborated by investigations of models similar to the present one which demonstrate the emergence of scale-free, system spanning correlations in the internal stress and local strain patterns \cite{Zaiser2006,Kapetanou2015}. As a consequence of such correlations, the emergent macroscopic materials behavior can neither be inferred from local statistics (e.g. using weakest-link arguments) nor can it be related to the properties of small, circumscribed representative volume element. Thus, novel conceptual tools may be needed to exploit the possibilities created for improving materials performance by exploiting the manner in which local fluctuations in materials properties may not only influence, but qualitatively change the macroscopic behavior of materials. 

\section*{Acknowledgements}

Financial support of the Hungarian Scientific Research Fund (OTKA) under
contract number PD-105256 and of the European Commission under grant agreement No.
CIG-321842 are also acknowledged. PDI is also supported by the J\'anos Bolyai Scholarship of the
Hungarian Academy of Sciences.

\bibliography{references}

\begin{thebibliography}{18}%
\makeatletter
\providecommand \@ifxundefined [1]{%
 \@ifx{#1\undefined}
}%
\providecommand \@ifnum [1]{%
 \ifnum #1\expandafter \@firstoftwo
 \else \expandafter \@secondoftwo
 \fi
}%
\providecommand \@ifx [1]{%
 \ifx #1\expandafter \@firstoftwo
 \else \expandafter \@secondoftwo
 \fi
}%
\providecommand \natexlab [1]{#1}%
\providecommand \enquote  [1]{``#1''}%
\providecommand \bibnamefont  [1]{#1}%
\providecommand \bibfnamefont [1]{#1}%
\providecommand \citenamefont [1]{#1}%
\providecommand \href@noop [0]{\@secondoftwo}%
\providecommand \href [0]{\begingroup \@sanitize@url \@href}%
\providecommand \@href[1]{\@@startlink{#1}\@@href}%
\providecommand \@@href[1]{\endgroup#1\@@endlink}%
\providecommand \@sanitize@url [0]{\catcode `\\12\catcode `\$12\catcode
  `\&12\catcode `\#12\catcode `\^12\catcode `\_12\catcode `\%12\relax}%
\providecommand \@@startlink[1]{}%
\providecommand \@@endlink[0]{}%
\providecommand \url  [0]{\begingroup\@sanitize@url \@url }%
\providecommand \@url [1]{\endgroup\@href {#1}{\urlprefix }}%
\providecommand \urlprefix  [0]{URL }%
\providecommand \Eprint [0]{\href }%
\providecommand \doibase [0]{http://dx.doi.org/}%
\providecommand \selectlanguage [0]{\@gobble}%
\providecommand \bibinfo  [0]{\@secondoftwo}%
\providecommand \bibfield  [0]{\@secondoftwo}%
\providecommand \translation [1]{[#1]}%
\providecommand \BibitemOpen [0]{}%
\providecommand \bibitemStop [0]{}%
\providecommand \bibitemNoStop [0]{.\EOS\space}%
\providecommand \EOS [0]{\spacefactor3000\relax}%
\providecommand \BibitemShut  [1]{\csname bibitem#1\endcsname}%
\let\auto@bib@innerbib\@empty
\bibitem [{\citenamefont {Ashby}\ and\ \citenamefont
  {Greer}(2006)}]{Ashby2006}%
  \BibitemOpen
  \bibfield  {author} {\bibinfo {author} {\bibfnamefont {M.}~\bibnamefont
  {Ashby}}\ and\ \bibinfo {author} {\bibfnamefont {A.}~\bibnamefont {Greer}},\
  }\href {\doibase http://dx.doi.org/10.1016/j.scriptamat.2005.09.051}
  {\bibfield  {journal} {\bibinfo  {journal} {Scripta Materialia}\ }\textbf
  {\bibinfo {volume} {54}},\ \bibinfo {pages} {321 } (\bibinfo {year}
  {2006})}\BibitemShut {NoStop}%
\bibitem [{\citenamefont {Steif}\ \emph {et~al.}(1982)\citenamefont {Steif},
  \citenamefont {Spaepen},\ and\ \citenamefont {Hutchinson}}]{Steif1982}%
  \BibitemOpen
  \bibfield  {author} {\bibinfo {author} {\bibfnamefont {P.}~\bibnamefont
  {Steif}}, \bibinfo {author} {\bibfnamefont {F.}~\bibnamefont {Spaepen}}, \
  and\ \bibinfo {author} {\bibfnamefont {J.~W.}\ \bibnamefont {Hutchinson}},\
  }\href@noop {} {\bibfield  {journal} {\bibinfo  {journal} {Acta
  Metallurgica}\ }\textbf {\bibinfo {volume} {30}},\ \bibinfo {pages} {447}
  (\bibinfo {year} {1982})}\BibitemShut {NoStop}%
\bibitem [{\citenamefont {Wright}\ \emph {et~al.}(2001)\citenamefont {Wright},
  \citenamefont {Schwarz},\ and\ \citenamefont {Nix}}]{Wright2001}%
  \BibitemOpen
  \bibfield  {author} {\bibinfo {author} {\bibfnamefont {W.}~\bibnamefont
  {Wright}}, \bibinfo {author} {\bibfnamefont {R.~B.}\ \bibnamefont {Schwarz}},
  \ and\ \bibinfo {author} {\bibfnamefont {W.}~\bibnamefont {Nix}},\
  }\href@noop {} {\bibfield  {journal} {\bibinfo  {journal} {Materials Science
  and Engineering A}\ }\textbf {\bibinfo {volume} {319}},\ \bibinfo {pages}
  {229} (\bibinfo {year} {2001})}\BibitemShut {NoStop}%
\bibitem [{\citenamefont {Zaiser}\ \emph {et~al.}(2013)\citenamefont {Zaiser},
  \citenamefont {Mill}, \citenamefont {Konstantinidis},\ and\ \citenamefont
  {Aifantis}}]{Zaiser2013}%
  \BibitemOpen
  \bibfield  {author} {\bibinfo {author} {\bibfnamefont {M.}~\bibnamefont
  {Zaiser}}, \bibinfo {author} {\bibfnamefont {F.}~\bibnamefont {Mill}},
  \bibinfo {author} {\bibfnamefont {A.}~\bibnamefont {Konstantinidis}}, \ and\
  \bibinfo {author} {\bibfnamefont {K.}~\bibnamefont {Aifantis}},\ }\href@noop
  {} {\bibfield  {journal} {\bibinfo  {journal} {Materials Science and
  Engineering A}\ }\textbf {\bibinfo {volume} {567}},\ \bibinfo {pages} {38}
  (\bibinfo {year} {2013})}\BibitemShut {NoStop}%
\bibitem [{\citenamefont {Zaiser}\ and\ \citenamefont
  {Moretti}(2005)}]{Zaiser2005}%
  \BibitemOpen
  \bibfield  {author} {\bibinfo {author} {\bibfnamefont {M.}~\bibnamefont
  {Zaiser}}\ and\ \bibinfo {author} {\bibfnamefont {P.}~\bibnamefont
  {Moretti}},\ }\href {http://stacks.iop.org/1742-5468/2005/i=08/a=P08004}
  {\bibfield  {journal} {\bibinfo  {journal} {Journal of Statistical Mechanics:
  Theory and Experiment}\ }\textbf {\bibinfo {volume} {2005}},\ \bibinfo
  {pages} {P08004} (\bibinfo {year} {2005})}\BibitemShut {NoStop}%
\bibitem [{\citenamefont {Talamali}\ \emph {et~al.}(2012)\citenamefont
  {Talamali}, \citenamefont {Petäjä}, \citenamefont {Vandembroucq},\ and\
  \citenamefont {Roux}}]{Talamali2012}%
  \BibitemOpen
  \bibfield  {author} {\bibinfo {author} {\bibfnamefont {M.}~\bibnamefont
  {Talamali}}, \bibinfo {author} {\bibfnamefont {V.}~\bibnamefont {Petäjä}},
  \bibinfo {author} {\bibfnamefont {D.}~\bibnamefont {Vandembroucq}}, \ and\
  \bibinfo {author} {\bibfnamefont {S.}~\bibnamefont {Roux}},\ }\href {\doibase
  http://dx.doi.org/10.1016/j.crme.2012.02.010} {\bibfield  {journal} {\bibinfo
   {journal} {Comptes Rendus Mécanique}\ }\textbf {\bibinfo {volume} {340}},\
  \bibinfo {pages} {275 } (\bibinfo {year} {2012})},\ \bibinfo {note} {recent
  Advances in Micromechanics of Materials}\BibitemShut {NoStop}%
\bibitem [{\citenamefont {Budrikis}\ and\ \citenamefont
  {Zapperi}(2013)}]{Budrikis2013}%
  \BibitemOpen
  \bibfield  {author} {\bibinfo {author} {\bibfnamefont {Z.}~\bibnamefont
  {Budrikis}}\ and\ \bibinfo {author} {\bibfnamefont {S.}~\bibnamefont
  {Zapperi}},\ }\href {\doibase 10.1103/PhysRevE.88.062403} {\bibfield
  {journal} {\bibinfo  {journal} {Phys. Rev. E}\ }\textbf {\bibinfo {volume}
  {88}},\ \bibinfo {pages} {062403} (\bibinfo {year} {2013})}\BibitemShut
  {NoStop}%
\bibitem [{\citenamefont {Sandfeld}\ \emph {et~al.}(2015)\citenamefont
  {Sandfeld}, \citenamefont {Budrikis}, \citenamefont {Zapperi},\ and\
  \citenamefont {Fernandez~Castellanos}}]{Sandfeld2015}%
  \BibitemOpen
  \bibfield  {author} {\bibinfo {author} {\bibfnamefont {S.}~\bibnamefont
  {Sandfeld}}, \bibinfo {author} {\bibfnamefont {Z.}~\bibnamefont {Budrikis}},
  \bibinfo {author} {\bibfnamefont {S.}~\bibnamefont {Zapperi}}, \ and\
  \bibinfo {author} {\bibfnamefont {D.}~\bibnamefont {Fernandez~Castellanos}},\
  }\href {http://stacks.iop.org/1742-5468/2015/i=2/a=P02011} {\bibfield
  {journal} {\bibinfo  {journal} {Journal of Statistical Mechanics: Theory and
  Experiment}\ }\textbf {\bibinfo {volume} {2015}},\ \bibinfo {pages} {P02011}
  (\bibinfo {year} {2015})}\BibitemShut {NoStop}%
\bibitem [{\citenamefont {Lin}\ \emph {et~al.}(2015)\citenamefont {Lin},
  \citenamefont {Gueudre}, \citenamefont {Rosso},\ and\ \citenamefont
  {Wyart}}]{Lin2015}%
  \BibitemOpen
  \bibfield  {author} {\bibinfo {author} {\bibfnamefont {J.}~\bibnamefont
  {Lin}}, \bibinfo {author} {\bibfnamefont {T.}~\bibnamefont {Gueudre}},
  \bibinfo {author} {\bibfnamefont {A.}~\bibnamefont {Rosso}}, \ and\ \bibinfo
  {author} {\bibfnamefont {M.}~\bibnamefont {Wyart}},\ }\href
  {http://link.aps.org/doi/10.1103/PhysRevLett.103.065501} {\bibfield
  {journal} {\bibinfo  {journal} {Phys. Rev. Lett.}\ }\textbf {\bibinfo
  {volume} {115}},\ \bibinfo {pages} {168001} (\bibinfo {year}
  {2015})}\BibitemShut {NoStop}%
\bibitem [{\citenamefont {Eshelby}(1957)}]{Eshelby1957}%
  \BibitemOpen
  \bibfield  {author} {\bibinfo {author} {\bibfnamefont {J.~D.}\ \bibnamefont
  {Eshelby}},\ }\href {\doibase 10.1098/rspa.1957.0133} {\bibfield  {journal}
  {\bibinfo  {journal} {Proceedings of the Royal Society of London A:
  Mathematical, Physical and Engineering Sciences}\ }\textbf {\bibinfo {volume}
  {241}},\ \bibinfo {pages} {376} (\bibinfo {year} {1957})}\BibitemShut
  {NoStop}%
\bibitem [{\citenamefont {Hirth}\ and\ \citenamefont
  {Lothe}(1982)}]{Hirth1982}%
  \BibitemOpen
  \bibfield  {author} {\bibinfo {author} {\bibfnamefont {J.}~\bibnamefont
  {Hirth}}\ and\ \bibinfo {author} {\bibfnamefont {J.}~\bibnamefont {Lothe}},\
  }\href {https://books.google.de/books?id=LFZGAAAAYAAJ} {\emph {\bibinfo
  {title} {Theory of Dislocations}}}\ (\bibinfo  {publisher} {Krieger
  Publishing Company},\ \bibinfo {year} {1982})\ pp.\ \bibinfo {pages}
  {59--78}\BibitemShut {NoStop}%
\bibitem [{\citenamefont {Bako}\ \emph {et~al.}(2006)\citenamefont {Bako},
  \citenamefont {Groma}, \citenamefont {Gy\"orgyi},\ and\ \citenamefont
  {Zimanyi}}]{Bako2006}%
  \BibitemOpen
  \bibfield  {author} {\bibinfo {author} {\bibfnamefont {B.}~\bibnamefont
  {Bako}}, \bibinfo {author} {\bibfnamefont {I.}~\bibnamefont {Groma}},
  \bibinfo {author} {\bibfnamefont {G.}~\bibnamefont {Gy\"orgyi}}, \ and\
  \bibinfo {author} {\bibfnamefont {G.}~\bibnamefont {Zimanyi}},\ }\href@noop
  {} {\bibfield  {journal} {\bibinfo  {journal} {Computational Materials
  Science}\ }\textbf {\bibinfo {volume} {38}},\ \bibinfo {pages} {22} (\bibinfo
  {year} {2006})}\BibitemShut {NoStop}%
\bibitem [{\citenamefont {Lin}\ \emph {et~al.}(2014)\citenamefont {Lin},
  \citenamefont {Saade}, \citenamefont {Lerner}, \citenamefont {Rosso},\ and\
  \citenamefont {Wyart}}]{Lin2014}%
  \BibitemOpen
  \bibfield  {author} {\bibinfo {author} {\bibfnamefont {J.}~\bibnamefont
  {Lin}}, \bibinfo {author} {\bibfnamefont {A.}~\bibnamefont {Saade}}, \bibinfo
  {author} {\bibfnamefont {E.}~\bibnamefont {Lerner}}, \bibinfo {author}
  {\bibfnamefont {A.}~\bibnamefont {Rosso}}, \ and\ \bibinfo {author}
  {\bibfnamefont {M.}~\bibnamefont {Wyart}},\ }\href@noop {} {\bibfield
  {journal} {\bibinfo  {journal} {Europhysics Letters}\ }\textbf {\bibinfo
  {volume} {105}},\ \bibinfo {pages} {26003} (\bibinfo {year}
  {2014})}\BibitemShut {NoStop}%
\bibitem [{\citenamefont {Ispánovity}\ \emph {et~al.}(2013)\citenamefont
  {Ispánovity}, \citenamefont {Hegyi}, \citenamefont {Groma}, \citenamefont
  {G.Györgyi}, \citenamefont {K.Ratter},\ and\ \citenamefont
  {Weygand}}]{Ispanovity2013}%
  \BibitemOpen
  \bibfield  {author} {\bibinfo {author} {\bibfnamefont {P.}~\bibnamefont
  {Ispánovity}}, \bibinfo {author} {\bibfnamefont {A.}~\bibnamefont {Hegyi}},
  \bibinfo {author} {\bibfnamefont {I.}~\bibnamefont {Groma}}, \bibinfo
  {author} {\bibnamefont {G.Györgyi}}, \bibinfo {author} {\bibnamefont
  {K.Ratter}}, \ and\ \bibinfo {author} {\bibfnamefont {D.}~\bibnamefont
  {Weygand}},\ }\href@noop {} {\bibfield  {journal} {\bibinfo  {journal} {Acta
  Materialia}\ }\textbf {\bibinfo {volume} {61}},\ \bibinfo {pages} {6234}
  (\bibinfo {year} {2013})}\BibitemShut {NoStop}%
\bibitem [{\citenamefont {S.Adibi}\ \emph {et~al.}(2013)\citenamefont
  {S.Adibi}, \citenamefont {Sha}, \citenamefont {Branicio}, \citenamefont
  {S.P.Joshi}, \citenamefont {Liu},\ and\ \citenamefont {Zhang}}]{Adibi2013}%
  \BibitemOpen
  \bibfield  {author} {\bibinfo {author} {\bibnamefont {S.Adibi}}, \bibinfo
  {author} {\bibfnamefont {Z.}~\bibnamefont {Sha}}, \bibinfo {author}
  {\bibfnamefont {P.}~\bibnamefont {Branicio}}, \bibinfo {author} {\bibnamefont
  {S.P.Joshi}}, \bibinfo {author} {\bibfnamefont {Z.}~\bibnamefont {Liu}}, \
  and\ \bibinfo {author} {\bibfnamefont {Y.}~\bibnamefont {Zhang}},\
  }\href@noop {} {\bibfield  {journal} {\bibinfo  {journal} {Applied Physics
  Letters}\ }\textbf {\bibinfo {volume} {103}},\ \bibinfo {pages} {211905}
  (\bibinfo {year} {2013})}\BibitemShut {NoStop}%
\bibitem [{\citenamefont {Das}\ \emph {et~al.}(2005)\citenamefont {Das},
  \citenamefont {Tang}, \citenamefont {Kim}, \citenamefont {R.Theissmann},
  \citenamefont {Baier}, \citenamefont {W.H.Wang},\ and\ \citenamefont
  {Eckert}}]{Das2005}%
  \BibitemOpen
  \bibfield  {author} {\bibinfo {author} {\bibfnamefont {J.}~\bibnamefont
  {Das}}, \bibinfo {author} {\bibfnamefont {M.}~\bibnamefont {Tang}}, \bibinfo
  {author} {\bibfnamefont {K.~B.}\ \bibnamefont {Kim}}, \bibinfo {author}
  {\bibnamefont {R.Theissmann}}, \bibinfo {author} {\bibfnamefont
  {F.}~\bibnamefont {Baier}}, \bibinfo {author} {\bibnamefont {W.H.Wang}}, \
  and\ \bibinfo {author} {\bibfnamefont {J.}~\bibnamefont {Eckert}},\
  }\href@noop {} {\bibfield  {journal} {\bibinfo  {journal} {Physical Review
  Letters}\ }\textbf {\bibinfo {volume} {94}},\ \bibinfo {pages} {205501}
  (\bibinfo {year} {2005})}\BibitemShut {NoStop}%
\bibitem [{\citenamefont {Zaiser}(2006)}]{Zaiser2006}%
  \BibitemOpen
  \bibfield  {author} {\bibinfo {author} {\bibfnamefont {M.}~\bibnamefont
  {Zaiser}},\ }\href@noop {} {\bibfield  {journal} {\bibinfo  {journal}
  {Advances in physics}\ }\textbf {\bibinfo {volume} {55}},\ \bibinfo {pages}
  {185} (\bibinfo {year} {2006})}\BibitemShut {NoStop}%
\bibitem [{\citenamefont {Kapetanou}\ \emph {et~al.}(2015)\citenamefont
  {Kapetanou}, \citenamefont {D.Weygand},\ and\ \citenamefont
  {Zaiser}}]{Kapetanou2015}%
  \BibitemOpen
  \bibfield  {author} {\bibinfo {author} {\bibfnamefont {O.}~\bibnamefont
  {Kapetanou}}, \bibinfo {author} {\bibnamefont {D.Weygand}}, \ and\ \bibinfo
  {author} {\bibfnamefont {M.}~\bibnamefont {Zaiser}},\ }\href@noop {}
  {\bibfield  {journal} {\bibinfo  {journal} {Journal of Statistical Mechanics:
  Theory and Experiment}\ ,\ \bibinfo {pages} {P08009}} (\bibinfo {year}
  {2015})}\BibitemShut {NoStop}%
\end{thebibliography}%
\end{document}